\pdfoutput=1
\documentclass[aps,prl,10pt,twocolumn,notitlepage,superscriptaddress]{revtex4-2}
\usepackage[utf8]{inputenc}
\usepackage{amsmath,amssymb,amsfonts,graphicx,times,mathtools,amsthm,centernot}
\usepackage{bbm}
\usepackage{dsfont}
\usepackage{bbold}

\usepackage{bm}
\usepackage{braket}
\usepackage[dvipsnames]{xcolor}
\usepackage{empheq}
\usepackage[breaklinks=true,colorlinks=true,linkcolor=teal,urlcolor=teal,citecolor=teal]{hyperref}
\usepackage{orcidlink}

\newcommand{\id}{\mathbb{1}}

\newcommand{\R}{\mathbb{R}}

\newcommand{\I}{\mathrm{i}}

\let\Re\relax
\let\Im\relax

\DeclareMathOperator{\Re}{Re}
\DeclareMathOperator{\Im}{Im}
\DeclareMathOperator{\Tr}{Tr}
\DeclareMathOperator{\tr}{tr}

\renewcommand{\vec}{\boldsymbol}
\let\originalleft\left
\let\originalright\right
\renewcommand{\left}{\mathopen{}\mathclose\bgroup\originalleft}
\renewcommand{\right}{\aftergroup\egroup\originalright}

\newcommand{\rvec}{\hat{\bm r}}
\newcommand{\OmegaM}{\bm\Omega}
\newcommand{\tran}{\mathsf{T}}

\newcommand{\yb}{\bm\eta}
\newcommand{\zb}{\bm\zeta}

\newcommand{\bs}{\boldsymbol{s}}

\begin{document}
\title{
Efficient evaluation of fundamental sensitivity limits and full counting statistics \\
for continuously monitored Gaussian quantum systems}
\author{Francesco Albarelli\,\orcidlink{0000-0001-5775-168X}}
\email{francesco.albarelli@gmail.com}
\affiliation{Università di Parma, Dipartimento di Scienze Matematiche, Fisiche e Informatiche, I-43124 Parma, Italy}
\affiliation{INFN—Sezione di Milano-Bicocca, Gruppo Collegato di Parma, I-43124 Parma, Italy}

\author{Marco G. Genoni\,\orcidlink{0000-0001-7270-4742}}
\email{marco.genoni@unimi.it}
\affiliation{Department of Physics ``A. Pontremoli", Universit\`a degli Studi di Milano, I-20133 Milano, Italy}%

\date{\today}

\begin{abstract}
Generalized master equations (GMEs)---time-local but generally neither trace-preserving nor Hermiticity-preserving---are convenient tools to compute properties of the environment of an open or continuously monitored quantum system.
A two-sided master equation yields the fidelity and quantum Fisher information (QFI) of environment states, thereby setting fundamental limits for hypothesis testing and parameter estimation under continuous monitoring.
For unmonitored noise or inefficient detection, the QFI of the detectable part of the environment may be obtained from a recently derived GME acting on multiple system replicas.
Tilted master equations provide the full counting statistics of quantum jumps and diffusive measurements, enabling, e.g., studies of quantum thermodynamics beyond average values.
Here we focus on bosonic linear systems, governed by a quadratic Hamiltonian and linear jump operators, whose dynamics preserves Gaussianity.
For Gaussian initial operators, we recast a generic GME as a compact set of ordinary differential equations for the covariance matrix (a Riccati-type equation), first moments, and normalization.
These equations can be integrated efficiently without Hilbert-space truncation, and admit analytical results in simple settings.
We also provide specialized forms for fidelity/QFI and full counting statistics.
We illustrate the formalism with a continuously monitored optical parametric oscillator, using it to determine sensitivity limits for frequency estimation and to benchmark Hasegawa's thermodynamic uncertainty relations.
\end{abstract}

\maketitle

\emph{Introduction.}---Assessing and computing properties of the environment of open or continuously monitored quantum systems, going beyond the reduced dynamics, is a long standing problem with applications in quantum optics, thermodynamics, and metrology~\cite{Landi2024}.
In this context, a convenient approach is to recast environment-dependent properties as functionals of generalized system operators (generally not physical states) whose evolution is governed by generalized master equations (GMEs).
While this idea may be very general, in this letter we will mainly focus on classes of GMEs that have been applied in parameter estimation and hypothesis testing with continuously monitored systems~\cite{Gammelmark2014c,Molmer2015a,Macieszczak2016,Yang2023e,Landi2024,Yang2026a}, and full counting statistics~\cite{Zheng2003,Bagrets2003,Peng2007,Esposito2009,Garrahan2010,Schaller2014,Benito2016,Pigeon2016,Landi2024}.

For Gaussian systems of bosons with quadratic Hamiltonian and linear jump operators, see Eq.~\eqref{eq:def_HLGaussian}, the standard Lindblad master equation (ME) can be recast into linear ordinary differential equations (ODEs) for the dynamics of the covariance matrix and first moments~\cite{Wiseman2005,Genoni2016b,Serafini2023,Nurdin2017a}.
However, comparable moment equations for GMEs are not known in full generality, despite the relevance of the Gaussian description across several platforms, e.g. quantum optics, optomechanics, and atomic ensembles~\cite{Serafini2023}.
Specific calculations along these lines have appeared scattered across different fields: to evaluate the ultimate sensitivity limits for the estimation of parameters encoded by linear Hamiltonians~\cite{Albarelli2017a,Yokomizo2026},
and to study full counting statistics and thermodynamics of trajectories in harmonic oscillators~\cite{Pigeon2015,Pigeon2015a,Guarnieri2016,Brange2019b,Portugal2023,Kansanen2025}, and to describe superposition of Gaussian processes~\cite{Braccini2025a}.

In this letter we go beyond these particular instances and recast Gaussian GMEs, in full generality, as nonlinear ODEs for moments.
Building on this, we provide application-specific forms to evaluate the fundamental limits on parameter estimation and full counting statistics in continuously monitored Gaussian quantum systems.

In what follows, by GME we specifically mean a time-local first-order evolution equation for a generalized system operator $\hat{\mu}$, generally neither normalized nor Hermitian.
It is convenient to write a generic GME by splitting the contribution from commutators and anticommutators (square and curly braces, respectively), and from left--right multiplication:
\begin{eqnarray}
\label{eq:GMEmu}
    \dot{\hat{\mu}} =  \left\{ \hat{A} , \hat{\mu} \right\} 
    + \left[ \hat{C} , \hat{\mu} \right] + \sum_{k=1}^{N_{\text{J}}} \hat{L}_k \hat{\mu} \hat{R}_k^\dag ,
\end{eqnarray}
where the dot indicates a temporal derivative, and the temporal dependence is left implicit to simplify notation~\footnote{Time dependence is implicit not only for $\hat{\mu}$, but also the operators on the right-hand side may be time dependent.
However, Eq.~\eqref{eq:GMEmu} is time local, in contrast to integro-differential equations for open-system dynamics~\cite{Breuer2002,Vacchini2024}, also sometimes termed GMEs~\cite{Vacchini2016c}.}.
A generic embedding into Lindblad form using an auxiliary qubit is discussed in Ref.~\cite{Hush2015}.

\emph{System--environment interaction and continuous measurements.}---We briefly recall the standard formalism to describe continuously-monitored quantum systems, we refer to Refs.~\cite{Gardiner2004,Wiseman2010,Jacobs2014a,Albarelli2024} for details.
The environment is modeled as continuum families of bosonic fields, which we assume initially in the vacuum.
The total system-environment (SE) Hamiltonian in the rotating frame with respect to the free Hamiltonian of the fields is $ \hat{H}_{\mathrm{SE},\theta} = \hat{H}_{\theta} -\I \left(\sum_k \hat{J}_{\theta,k} \hat{a}_{k}^\dag(t)  - \text{h.c} \right)$, where $[\hat{a}_k(t),\hat{a}_j(t')^\dag ] = \delta_{kj} \delta(t-t')$ are quantum white-noise operators.
Preparing the ground for parameter estimation tasks, we have highlighted the dependence on an arbitrary parameter $\theta$.
Assuming a pure initial state for the system, the total SE state $\ket{\Psi_{\mathrm{SE},\theta}}$ evolves unitarily and remains pure, mathematically it is a continuous matrix product state~\cite{Verstraete2010,Hasegawa2023a}.
The reduced state of the environment $\rho_{\mathrm{E},\theta} = \Tr_{\mathrm{S}}[ | \Psi_{\mathrm{SE},\theta} \rangle \langle \Psi_{\mathrm{SE},\theta} | ]$ is mixed, a so-called continuous matrix product operator~\cite{Tjoa2025}.
On the other hand, tracing out the environment leads to a Lindblad ME for the system, which corresponds to the choice $\hat{C}=-\I \hat{H}_{\theta}$, $\hat{A} = -(1/2)\sum_k \hat{J}_{\theta,k}^\dag \hat{J}_{\theta,k}$ and $\hat{L}_k = \hat{R}_k = \hat{J}_{\theta,k}$ in Eq.~\eqref{eq:GMEmu}~\footnote{While this environment model can be used as a unitary dilation of any Lindblad ME, the same dynamics can be obtained from a different microscopic derivation, often assuming weak coupling with an environment initially in a thermal state~\cite{Breuer2002,Vacchini2024}.}.

Despite being labeled as \emph{environment}, the continuum families $\{ \hat{a}_k(t) \}$ can often be interpreted as various output channels for the field emitted by the system.
In principle, these fields can be measured, and often the system itself is actually inaccessible (e.g. the field inside a cavity).
When the modes $\hat{a}_k(t)$ are instantaneously measured, the system experiences a back-action and the dynamics is a \emph{quantum trajectory} governed by a stochastic master equation (SME) for the \emph{conditional} state, also known as an \emph{unravelling} of the Lindblad ME, which in turn represents the \emph{unconditional} dynamics and it is equivalent to averaging over trajectories.
The experimental observation of individual quantum trajectories in continuously monitored quantum systems has now been demonstrated across several platforms, including superconducting circuits~\cite{Murch2013,Campagne-Ibarcq2016,Ficheux2017,Minev2019}, optomechanical systems~\cite{Wieczorek2015,Rossi2018,Mason2018}, and hybrid quantum architectures~\cite{Thomas2020}.

\emph{Continuous variables.}---We consider a system of $n$ bosonic modes.
Let $\rvec := (\hat x_1,\hat p_1,\dots,\hat x_n,\hat p_n)^{\mathsf T}$ be a vector of canonical operators satisfying $[\hat r_j,\hat r_k]=\I \,\Omega_{jk}$, where $\Omega := \bigoplus_{j=1}^n\begin{pmatrix}0&1\\-1&0\end{pmatrix}$.
For any trace-class operator $\hat{O}$, we define the symmetrically-ordered (Weyl) characteristic function
$\chi_{\hat{O}}(\bs):=\Tr\left[\hat{O}\,\hat{D}(-\bs)\right]$ as the expectation value of the displacement operators $\hat{D}(\bs):=\exp\! \left[\I \,\bs^\tran\OmegaM \hat{\vec{r}} \right]$, where $\bs \in \R^{2n}$ is a phase-space coordinate, and it is often convenient to use the variable $\tilde{\vec{r}}=\Omega \bs$.
For a normalized operator $\hat{\nu}$ with $\Tr[\hat{\nu}] = 1$, we define first moments vector and covariance matrix (CM) as
\begin{equation}
\vec{d} := \Tr \left[ \rvec \hat{\nu} \right],\; \;
\sigma:=\Tr \left[\{\Delta\rvec,\Delta\rvec^{\mathsf T}\}\hat{\nu} \right],\; \;
\Delta\rvec:=\rvec-\vec{d},
\label{eq:serafini_cov}
\end{equation}
notice that in this convention the CM of the vacuum is $\id_{2n}$.
We call $\hat{\nu}$ a Gaussian operator if $\chi_{\hat{\nu}}$ is a Gaussian function, such that it is completely specified by $\vec{d}$ and $\sigma$.
If $\hat{\nu}$ is positive semidefinite and unit-trace it is a standard Gaussian state~\cite{Serafini2023}.

The Gaussianity of states is preserved for a physical dynamics with quadratic Hamiltonians and linear jump operators:
\begin{equation}
\label{eq:def_HLGaussian}
\hat{H}_{\theta} = \frac12 \hat{\vec{r}}^\tran \mathbb{H}_{\theta} \hat{\vec{r}} + \mathbb{h}_{\theta}^\tran \Omega \hat{\vec{r}} , \quad \hat{\vec{J}}_{\theta} = \mathbb{J}_{\theta} \hat{\vec{r}} \, , 
\end{equation}
with $\mathbb{H}_{\theta} = \mathbb{H}_{\theta}^\tran \in \mathbbm{R}^{2n{\times}2n}$, $\mathbb{h}_{\theta} \in \mathbbm{R}^{2n}$ since $\hat{H}_{\theta}$ must be Hermitian, and $\mathbb{J}_{\theta} \in \mathbbm{C}^{N_{\mathrm{J}}{\times} 2n} $.
Similarly, the Gaussianity of an operator evolved with GME in Eq.~\eqref{eq:GMEmu} is preserved if the superoperators on the right-hand side are quadratic:
\begin{align}
    \hat{A} = \frac12 \hat{\vec{r}}^{\tran} \mathbb{A} \hat{\vec{r}} + \mathbb{a}^\tran \Omega \hat{\vec{r}}  & \qquad  \hat{C} = \frac12 \hat{\vec{r}}^{\tran} \mathbb{C} \hat{\vec{r}} + \mathbb{c}^\tran \Omega \hat{\vec{r}} \label{eq:quadratic-A-C}\\ 
    \sum_{k}^{N_{\text{J}}} \hat{L}_k \hat{\mu} \hat{R}_k^\dag = &  \sum_{mn} \mathbb{K}_{mn} \hat{r}_m \hat{\mu} \hat{r}_n \, , \label{eq:LmuR_linear}
\end{align}
where now $\mathbb{A}, \mathbb{C} \in \mathbbm{C}^{2n{\times}2n}$ as $\hat{A}$ and $\hat{C}$ need not be Hermitian operators.
The superoperator in Eq.~\eqref{eq:LmuR_linear} corresponds to $\hat{\vec{L}}  = \mathbb{L} \hat{\vec{r}}, \;  \hat{\vec{R}} = \mathbb{R}  \hat{\vec{r}}$, where $\mathbb{K} = (\mathbb{R}^\dag \mathbb{L})^\tran$.

\emph{Generalized moment equations.}---First, it is convenient to work with a normalized operator $\hat{\nu}$, so we introduce a complex scalar $\xi$ such that $\hat{\mu} = e^{-\xi} \hat{\nu} $ and $\Tr[ \hat{\mu}] = e^{-\xi}$.
Second, we parametrize the normalized Gaussian operator $\hat{\nu}$ as:
\begin{equation}
\label{eq:nu_Gauss_charfun}
\chi_{\hat{\nu}}(\tilde{\boldsymbol{r}}) =
\exp\!\left[-\frac14\,\tilde{\boldsymbol{r}}^{\mathsf T} \sigma \,\tilde{\boldsymbol{r}} \right]\,
\exp\!\left[\I \,\tilde{\boldsymbol{r}}^{\tran} \vec{d} \right].
\end{equation}
Notice that $\sigma \in \mathbbm{C}^{2n{\times}2n}$ and $\vec{d}\in \mathbbm{C}^{2n}$ have complex coefficients.
Thanks to the symmetry of the quadratic form, we can assume a symmetric covariance matrix $\sigma^\tran = \sigma$ without loss of generality.

The action of canonical operators can be mapped to differential operators acting on the characteristic function~\cite{Gardiner2004,Serafini2023}.
Thus, by computing the characteristic function of both sides of Eq.~\eqref{eq:GMEmu} with the parametrization in Eq.~\eqref{eq:nu_Gauss_charfun}, one obtains quadratic polynomials in $\tilde{\vec{r}}$ multiplying the function $\chi_{\hat{\nu}}(\tilde{\vec{r}})$, the explicit correspondence is reported in the End Matter.
For the equation to hold at the operator level, the coefficients of the two polynomials must be equal.
In this way, with lengthy but straightforward calculations, we obtain a set of coupled ODEs:
\begin{align}
\dot{\sigma} & = \sigma Q \sigma + A  \sigma + \sigma A^\tran + D 
\label{eq:dotsigma}
\\ 
 \dot{\vec{d}} & = A \vec{d} + \sigma Q\vec{d} - \sigma \Omega \mathbb{a} + \I \mathbb{c} \label{eq:dotd} \\
 \dot{\xi} & = - \vec{d}^\tran Q \vec{d} - 
 \frac{1}{2} \tr \left[  
 \sigma  Q  \right] - \frac{1}{2} \tr \left[ W  \Omega \right]  - 2 \mathbb{a}^\tran \Omega \vec{d} \,  \label{eq:doteta}
\end{align}
where the coefficient matrices are related to the original operators as
\begin{align}
     A & =  \Omega \left( \I \operatorname{Sym}[\mathbb{C}]  + \I \operatorname{Skew}[\mathbb{K}]  \right) , \; \; 
      D    =  \Omega  \operatorname{Sym}[ \mathbb{K}- \mathbb{A}]  \Omega^\tran  \nonumber \\
     Q   & =   \operatorname{Sym}[ \mathbb{K} + \mathbb{A}] , \; \; 
     W  =  \I \operatorname{Skew}[ \mathbb{K}- \mathbb{A}] \, ,
     \label{eq:AandDdef}
\end{align}
where $\operatorname{Sym}[M] = (M + M^\tran )/2$ and $\operatorname{Skew}[M] = (M - M^\tran )/2$.
This set of ODEs is the central result of this Letter.
The standard Lindblad ME for Gaussian dynamics with the Hamiltonian and jump operators in Eq.~\eqref{eq:def_HLGaussian} corresponds to $\mathbb{C} = -\I \mathbb{H}$, $\mathbb{c} = - \I \mathbb{h}$,  $\mathbb{a} = 0$, and $\mathbb{L} = \mathbb{R} = \mathbb{J}$, which means $\mathbb{K}^\tran = \mathbb{J}^\dag \mathbb{J} = -\mathbb{A}$, eventually leading to the well-known drift and diffusion moment equations~\footnote{$\dot{\sigma} = A \sigma + \sigma A^\tran + D$ and $\dot{\vec{d}} = A \vec{d} + \mathbb{h}$, with $A=\Omega \left( \mathbb{H} + \Im[\mathbb{J}^\dag \mathbb{J} ] \right)$ and $D = 2 \Omega \Re[ \mathbb{J}^\dag \mathbb{J} ]  \Omega^\tran$,  cf.~\cite[Eqs.~(10--11)]{Wiseman2005} (we have an additional factor 2 in $D$, due to a different definition of the covariance matrix).}, and $\dot{\xi} = 0$ as expected for CPTP dynamics.

We also note that the initial Gaussian operator need not be a physical state, for example the outer product of two pure Gaussian states is a Gaussian operator.
Thus, by linearity, Eqs.~\eqref{eq:dotsigma}--\eqref{eq:doteta} can be used to propagate finite sums of Gaussian operators, e.g. superpositions of pure Gaussian states, as done for regular MEs~\cite{Corney2003a,Bourassa2021,Braccini2025a}.

\emph{GME for system--environment overlap and environment fidelity.}---The overlap between two joint SE states evolving with different Hamiltonians $\hat{H}_{\mathrm{SE},\theta_1}$ and $\hat{H}_{\mathrm{SE},\theta_2}$ is obtained as $\langle \Psi_{\mathrm{SE},\theta_2}(t) |  \Psi_{\mathrm{SE},\theta_1}(t) \rangle = \Tr[\hat{\mu}(t)]$, where  the generalized density operator is initialized in a pure state of the system $ \hat{\mu}(0) = |\psi_0\rangle \langle \psi_0|$ and then evolved with the two-sided master equation (TSME)~\cite{Gammelmark2014c}:
\begin{align}
\label{eq:TSME_HS}
  \dot{\hat{\mu}} & = -\I \left( \hat{H}_{\theta_1} \hat{\mu} - \hat{\mu} \hat{H}_{\theta_2}  \right) + \sum_l   \hat{J}_{\theta_1,l} \; \hat{\mu} \; \hat{J}_{\theta_2,l}^\dag  \\  
  & \qquad -  \frac12 \left( \sum_l \hat{J}_{\theta_1,l}^\dag  \hat{J}_{\theta_1,l} \right) \hat{\mu} 
    -  \frac12 \hat{\mu}  \left( \sum_l \hat{J}_{\theta_2,l}^\dag \hat{J}_{\theta_2,l} \right) \, , \nonumber
\end{align}
which can easily be recast in the form of Eq.~\eqref{eq:GMEmu}.
The fidelity between two environment-only states for different values of the parameters can also be obtained from the operator $\hat{\mu}$, by computing the trace norm: $\mathcal{F}[\rho_{\mathrm{E},\theta_1},\rho_{\mathrm{E},\theta_2}] := \Tr\left[\sqrt{\sqrt{\rho_{\mathrm{E},\theta_1}}\rho_{\mathrm{E},\theta_2}\sqrt{\rho_{\mathrm{E},\theta_1}}}\right] = \Tr \left[\sqrt{\hat{\mu}^\dag \hat{\mu}} \right] = \left\Vert \hat{\mu} \right\Vert_1$~\cite{Yang2023e}.

These quantities set the ultimate limits for parameter estimation~\cite{Gammelmark2014c,Yang2023e} and hypothesis testing~\cite{Molmer2015a} with continuously-monitored quantum systems, and more generally when the emitted field can be measured, e.g., in spectroscopic scenarios~\cite{Khan2025,Chinni2025}.
For hypothesis testing, the optimal probability of error to distinguish two states (equally likely, a priori) is upper and lower bounded by the fidelity as $ \frac12 (1 - \sqrt{1-\mathcal{F}^2}) \leq p_{\mathrm{err}} \leq  \mathcal{F} / 2$; the lower bound is tight for pure states, where the fidelity is the absolute value of the overlap.

\begin{figure}
\includegraphics[width=\columnwidth]{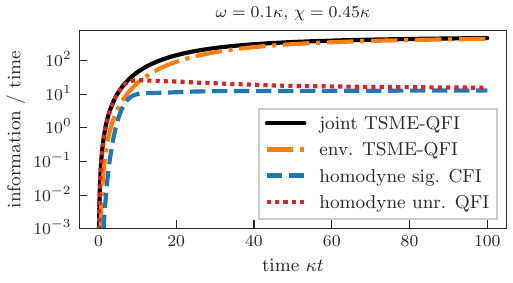}
\caption{Fisher information rates for $\omega$ estimation with an optical parametric oscillator (OPO), as a function of time.
The signal CFI and unravelling QFI are obtained for continuous homodyne detection.}
\label{fig1}
\end{figure}

The optimal sensitivity for the estimation of a parameter $\theta$ encoded in a quantum state $\rho_{\theta}$ is given by the quantum Fisher information (QFI)~\cite{Paris2009,Demkowicz-Dobrzanski2015a}, which is related to the second derivative of the fidelity: $ \sup_{\hat{O}} (\partial_{\theta} \Tr[\rho_{\theta} \hat{O}])^2 / \operatorname{Var}[\hat{O}] = \mathcal{Q}[ \rho_{\theta} ] := -4 \partial_\varepsilon^2 \log \mathcal{F}[ \rho_{\theta},\rho_{\theta+\varepsilon}]|_{\varepsilon=0}$ (the derivative can also be approximated numerically in practical scenarios).
More specifically, the joint SE QFI $\mathcal{Q}[ \ket{\Psi_{\mathrm{SE},\theta}} ]$ can be calculated from the pure-state fidelity obtained from the TSME as $\left| \Tr[ \hat{\mu} ] \right|$, which we dub \emph{joint} TSME-QFI.
Similarly, the environment-only QFI  $\mathcal{Q}[ \rho_{\mathrm{E},\theta} ]$ is obtained from the environmental fidelity $\left \Vert \hat{\mu} \right \Vert_1$, and we dub it \emph{environment} TSME-QFI.
Clearly, they satisfy $\mathcal{Q}[ \ket{\Psi_{\mathrm{SE},\theta}} ] \geq \mathcal{Q}[ \rho_{\mathrm{E},\theta} ] $.
Beyond parameter estimation, the joint TSME-QFI has been used to derive quantum thermodynamic uncertainty relations (TURs)~\cite{Hasegawa2020,Landi2024,Hasegawa2021a,VanVu2022a,Hasegawa2023a,Prech2025}.

We note that the TSME approach relies on the assumption of a pure joint SE state.
In the presence of unmonitored jump operators, evaluating the QFI becomes significantly harder~\cite{Khan2025}.
Ref.~\cite{Yang2026a} recently proposed an approach based on a GME acting on multiple system replicas, which we briefly review and adapt to the Gaussian case in the End Matter.

For a continuously monitored system we denote as $\mathcal{C}_{\mathrm{sig},\theta}$ the classical Fisher information (CFI) of the observed signal (e.g. photocurrent), obtained by performing a specific continuous measurement.
Roughly speaking, the signal CFI sets the fundamental limit on the parameter sensitivity when data is collected for a long time and can be processed with any estimator, taking into account all the temporal correlations in the signal.
We refer to Ref.~\cite{Gammelmark2013a} for its definition and more details about its evaluation~\cite{Albarelli2018c,Genoni2017}.
Since the signal CFI assumes a specific measurement of the fields, it is upper bounded by the environment TSME-QFI: $\mathcal{Q}[ \rho_{\mathrm{E},\theta} ] \geq \mathcal{C}_{\mathrm{sig},\theta}$.

After continuously monitoring the system for a certain time, in principle one can also perform a final measurement on the conditional state of the system $\rho_{\mathrm{S},\theta}^{\mathrm{(cond)}} $, i.e. a specific quantum trajectory obtained from a SME.
The information obtained in this scenario is still upper bounded by the joint TSME-QFI, and an appropriate figure of merit is the \emph{unravelling} QFI~\cite{Albarelli2017a,Catana2014} 
$\mathcal{Q}_{\mathrm{unr},\theta} := \mathcal{C}_{\mathrm{sig},\theta} + \mathbbm{E}_{\mathrm{traj}}\left[ \mathcal{Q}[\rho_{\mathrm{S},\theta}^{\mathrm{(cond)}}]  \right]  \leq \mathcal{Q}[ \ket{\Psi}_{\mathrm{SE},\theta} ] $, where the QFI of the conditional states is averaged over all possible trajectories (we refer to~\cite{Genoni2017,Zhang2025Backaction} and to the Supplemental Material~\cite{supp} for more details about the evaluation of such quantities in Gaussian systems).
For short times, the information can reside mostly in the conditional states, and often $\mathcal{Q}_{\mathrm{unr},\theta} \geq \mathcal{Q}[ \rho_{\mathrm{E},\theta} ]$.
At long times, for time-independent unconditional dynamics relaxing to a unique steady state, the environment and joint SE QFIs grow linearly in time~\cite{Gammelmark2014c,Macieszczak2016} with the same asymptotic rate; access to the system provides only an asymptotically constant amount of additional information~\cite{Yang2023e}.

For Gaussian systems, we obtain a specialized version of the general equations~\eqref{eq:dotsigma}--\eqref{eq:doteta} that correspond to the TSME in Eq.~\eqref{eq:TSME_HS}.
We introduce the notation $ \mathbb{H}^{(\pm)} := (\mathbb{H}_{\theta_1} \pm \mathbb{H}_{\theta_2} ) / 2 $, $\mathbb{G}^{(\pm)} := ( \mathbb{J}^\dag_{\theta_1}\mathbb{J}_{\theta_1} \pm \mathbb{J}^\dag_{\theta_2} \mathbb{J}_{\theta_2} )/2$,
and we have $\mathbb{K} = (\mathbb{J}_{\theta_2}^\dag \mathbb{J}_{\theta_1})^\tran $, where the objects on the right-hand side refer to the Hamiltonian and jump operators in Eq.~\eqref{eq:def_HLGaussian}.
This leads to the following matrices and vectors for the moment equations~\eqref{eq:dotsigma}--\eqref{eq:doteta}
\begin{align}
A_{\theta_1,\theta_2}
& =
\Omega\left[
\mathbb{H}^{(+)}
- \I \, \operatorname{Sym}[\mathbb{G}^{(-)}] + \I  \,\operatorname{Skew}\left[\mathbb{K}\right]
\right] \label{eq:A12}\\
D_{\theta_1,\theta_2}
& = 
\Omega\,
\operatorname{Sym}\!\left[
\mathbb{K}
+ \I \mathbb{H}^{(-)}
+ \mathbb{G}^{(+)}
\right]\,
\Omega^{\tran} \\
Q_{\theta_1,\theta_2}
&=
\operatorname{Sym}\!\left[
\mathbb{K}
- \I \mathbb{H}^{(-)}
-\mathbb{G}^{(+)}
\right] \\ 
W_{\theta_1,\theta_2}
& =
\I \,\operatorname{Skew}\!\left[
\mathbb{K}
+ \I \mathbb{H}^{(-)} + 
\mathbb{G}^{(+)}
\right] \\
\mathbb{a}_{\theta_1,\theta_2}
=& - \frac{\I}{2} \left( \mathbb{h}_{\theta_1} - \mathbb{h}_{\theta_2} \right), \;\; 
\mathbb{c}_{\theta_1,\theta_2}
= -\frac{\I}{2} \left( \mathbb{h}_{\theta_1} + \mathbb{h}_{\theta_2} \right) .
\label{eq:a_c_12}
\end{align}
Since the fidelity between joint SE states is $ |\langle \Psi_{\mathrm{SE},\theta_2}(t) |  \Psi_{\mathrm{SE},\theta_1}(t) \rangle | =|\Tr[ \hat{\mu} ]|  = e^{-\Re[\xi]}$, we obtain
\begin{equation}
\mathcal{Q} \left[ \ket{\Psi_{\mathrm{SE},\theta}(t)}\right] = 4 \Re\left[ \partial_\varepsilon^2 \xi (\theta, \theta+\varepsilon) \bigl|_{\varepsilon=0} \right] \,, \label{eq:QFIglobalGauss}
\end{equation}
where we have explicitly introduced the dependence of $\xi$ on $\theta_1=\theta$ and $\theta_2=\theta+\varepsilon$.
In the long-time limit, and in absence of multiple steady states~\cite{Macieszczak2016,Gammelmark2014c}, $\mathcal{Q} \left[ \ket{\Psi_{\mathrm{SE},\theta}(t)}\right]$ grows linearly in time and it is useful to introduce the steady-state QFI rate 
\begin{equation}
\lim_{t \to \infty} \frac{\mathcal{Q} \left[ \ket{\Psi_{\mathrm{SE},\theta}(t)} \right]}{t} = 4 \Re\left[ \partial_\varepsilon^2 \dot\xi_{\sf  ss}  (\theta, \theta+\varepsilon) \bigl|_{\varepsilon=0} \right] \,,
\label{eq:QFIglobalGaussrate}
\end{equation}
where $\dot\xi_{\sf  ss}$ corresponds to Eq.~\eqref{eq:doteta}, with covariance matrix $\sigma$ and first moments vector $\vec{d}$ replaced by the steady-state solutions of Eqs.~\eqref{eq:dotsigma} and~\eqref{eq:dotd}.\\
The environment-only fidelity reads
\begin{equation}
\label{eq:tracenorm_mu}
\begin{aligned}
   & \Tr \left[ \sqrt{\hat{\mu}^\dag \hat{\mu}} \right] =    e^{-\Re[\xi]} \frac{\exp\left( \Im[ \vec{d}]^\tran \Re[\sigma]^{-1} \Im[\vec{d}] \right)}{ \det( \Re[\sigma] )^{1/4} }  \\
   & \qquad \quad \cdot \left[ 2^{-n/2}\prod_{k=1}^n\Big(\sqrt{\nu_k+1}+\sqrt{\nu_k-1}\Big) \right] \, 
\end{aligned}
\end{equation}
where $\nu_k$ are the symplectic eigenvalues~\footnote{An $n$-mode Gaussian state has $n$ symplectic eigenvalues, obtained as the absolute values of the eigenvalues of $\I \Omega \sigma$ \cite{Serafini2023}. Its eigenvalues occur in opposite pairs because $\I\Omega\sigma$ is similar to $\I\sigma^{1/2}\Omega\sigma^{1/2}$, which is $\I$ times a real skew-symmetric matrix.} of the bona-fide CM $\frac12\left[ \Re[\sigma] + ( \Im[\sigma] + \Omega)\, \Re[\sigma]^{-1}\,(\Im[\sigma] + \Omega)^\tran \right]$; see Ref.~\cite{supp} for the explicit derivation.
While this formula does not lead to ready-to-use equations for the environment-only QFI  $\mathcal{Q}[ \rho_{\mathrm{E},\theta} ]$, it still allows to efficiently evaluate it via numerical methods.

We now consider the task of frequency-estimation with a paradigmatic Gaussian system, the optical parametric oscillator, i.e. a nonlinear crystal inside an optical cavity, described by the Hamiltonian $\hat{H}_{\sf OPO} = \omega \hat{a}^\dag \hat{a} + \I \frac{\chi}{2}\left( \hat{a}^2 - \hat{a}^{\dag 2} \right)$ and jump operator $\sqrt{\kappa} \hat{a}$, where we have defined the annihilation operator as $\hat{a} = (\hat{x} + \I \hat{p})/\sqrt{2}$~\cite{Genoni2016b,Fallani2022,Albarelli2024}.
We focus on the estimation of the Hamiltonian parameter $\omega$: for this model, the long-time joint TSME-QFI for $\omega = 0$ can be analytically evaluated via Eq.~\eqref{eq:QFIglobalGaussrate}, obtaining:
\begin{equation}
\lim_{t \to \infty} \frac{\mathcal{Q} \left[ \ket{\Psi_{\mathrm{SE},\omega}(t)} \right]}{t} = \frac{8 \kappa  \chi ^2 \left(5 \kappa ^2-4 \chi ^2\right)}{\left(\kappa ^2-4 \chi ^2\right)^3} ,
\label{eq:QFIrateOPO}
\end{equation}
showing a divergence for $\chi \to \kappa / 2$, which may be expected since the system becomes unstable, a feature that can be exploited for critical quantum metrology~\cite{Tsang2013, DiCandia2023,Fallani2022,Alushi2024,Blinova2026} (see Ref.~\cite{supp} for details on the derivation of Eq.~\eqref{eq:QFIrateOPO}).

Going beyond analytical steady-state values, we also computed finite-time quantities. 
The joint and environment TSME-QFIs are obtained by numerically solving the moment equations.
Even if these QFIs are obtained from the fidelity with a finite-difference approximation, the results show excellent stability even for long times, which may be harder to achieve with a Hilbert space truncation.
As a particular continuous monitoring strategy to estimate $\omega$, we consider homodyne detection of the light leaking out of the cavity. 
Since homodyne detection preserves Gaussianity, the signal CFI and unravelling QFI are also computed in terms of moments, exploiting the methods in Refs.~\cite{Genoni2017,Fallani2022}.

In Fig.~\ref{fig1} we show information rates as a function of time; one can see that the unravelling QFI for short time essentially attains the joint TSME-QFI, which signals that most of the information about the parameter is in the conditional states.
Since the average QFI of conditional states saturates to a constant value for sufficiently long times, the corresponding rate decreases and eventually the unravelling QFI tends to the signal CFI.
On the other hand, the signal CFI for homodyne detection never quite attains the environment TSME-QFI, with a gap that becomes very significant at long times.
This suggests that non-Gaussian measurements will be optimal for this task.
In the End Matter we present additional results for inefficient detection, obtained with the replica GME method.

\emph{GME for full counting statistics.}---We focus here on counting measurements, which in quantum optical scenarios correspond to monitoring each output field of the jump operators appearing in the Lindblad ME with photodetectors; diffusive measurements are discussed in the End Matter.
The goal of full counting statistics is to obtain the distribution of counts for a total charge $N(t) = \sum_k w_k N_k(t)$, a random variable where $w_k \in \mathbbm{R}$ are weights related to different jump operators, which may be used to express several quantities, e.g. energy or net amount of excitations leaving the system~\cite{Landi2024}.
It is convenient to express it as the Fourier transform of its characteristic function $P(n,t) = \int \frac{d \lambda}{2 \pi} e^{-\I n \lambda} \varphi(\lambda, t)$, which in turn can be obtained from a generalized density operator $\varphi(\lambda, t) = \Tr[\hat{\mu}]$~\footnote{We consider here the case of a single counting field $\lambda$, but the statistics of different quantities, such as work or heat exchanged with multiple environments, can be obtained from GMEs with suitable counting fields~\cite{Esposito2009,Soret2022,Kansanen2025}.}, where $\hat{\mu}$ is evolved with the tilted Lindblad ME~\cite{Esposito2009,Landi2024}
\begin{equation}
    \dot{\hat{\mu}} = -\I [ \hat{H} , \hat{\mu} ] + \sum_k \bigl( e^{\I \lambda w_k } \hat{J}_k \hat{\mu} \hat{J}_k^\dag - \frac{1}{2} \left\{ \hat{J}_k^\dag \hat{J}_k , \hat{\mu} \right\} \bigr) \, .
    \label{eq:tilted-ME-FCS}
\end{equation}

For Gaussian systems, the evolution given by Eq.~\eqref{eq:tilted-ME-FCS} only affects  the matrix $\mathbb{K}$ with respect to a standard physical Lindblad ME, leading to:
\begin{align}
\mathbb{K}_{\lambda} & = 
\left( \mathbb{J}^\dag \operatorname{diag}(e^{\I \lambda w_1 }, \dots , e^{\I \lambda w_{N_\mathrm{J}} }) \mathbb{J} \right)^\tran \, \label{eq:matrics_FCS}  \\
A_{\lambda} & =  \Omega \left( \mathbb{H} +  \I \operatorname{Skew}[\mathbb{K}_{\lambda}] \right),
\,\,\,
D_{\lambda}  =  \Omega \left( \operatorname{Sym}\left[ \mathbb{K}_{\lambda}\right]  + \Re \left[ \mathbb{J}^\dag \mathbb{J}  \right]
\right) \Omega^\tran \nonumber \\
Q_{\lambda} & = \operatorname{Sym}\left[ \mathbb{K}_{\lambda}\right] -\Re \left[ \mathbb{J}^\dag \mathbb{J}  \right], 
\,\,\,
W_\lambda = 
\I \operatorname{Skew}\left[ \mathbb{K}_{\lambda}\right] - \Im\left[\mathbb{J}^\dag \mathbb{J}\right], \;  \nonumber
\end{align}
and $\mathbb{a} = 0 \, , \; \mathbb{c} = -\I \mathbb{h}$.

\begin{figure}
\includegraphics{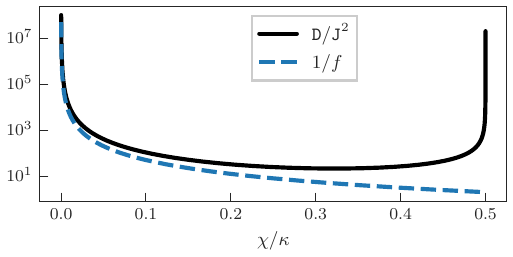}
\caption{The ratio $\mathtt{D}/\mathtt{J}^2$ and the right-hand side $1/f$ of Hasegawa's TUR for the OPO, as a function of $\chi / \kappa$, with $\omega = 0$.}
\label{fig2}
\end{figure}
Since the characteristic function here reads $\varphi(\lambda,t) = \Tr[\hat{\mu}]=e^{-\xi(\lambda,t)}$, it is often convenient to consider the cumulant generating function for jumps $C(\lambda,t) := \ln \varphi(\lambda,t) = -\xi(\lambda,t)$.
Also in this case, in the long-time limit, $C(\lambda,t)$ grows linearly in time and thus one introduces the scaled cumulant generating function (SCGF) $C(\lambda ) := \lim_{t\to \infty} \partial_t C(\lambda,t ) = - \dot{\xi}_{\sf  ss}(\lambda)$ from which one can readily compute the average current $\mathtt{J} := \lim_{t\to \infty} \frac{d}{dt} \mathbbm{E}[ N(t) ] = -\I \partial_{\lambda} C(\lambda) $ and the  noise $\mathtt{D} := \lim_{t\to \infty} \frac{d}{dt} \operatorname{Var}[N(t)] = ( -\I \partial_{\lambda})^2 C(\lambda) $, i.e. rate of change of the variance.
Hasegawa derived a TUR for these quantities~\cite{Hasegawa2020,Landi2024}:
$\mathtt{D} / \mathtt{J}^2 \geq 1 / f $, where $f = \lim_{t\to \infty} \mathcal{Q}[ \ket{\Psi_{\mathrm{SE},\theta}(t)} ]/t$ is the joint TSME-QFI rate for a parameter $\theta$ that deforms the dynamics as $\hat{H}_{\theta} = (1+\theta) \hat{H}$ and $\hat{J}_{\theta} = \sqrt{1 + \theta} \hat{J}$, and that can be evaluated as previously described via Eq.~\eqref{eq:QFIglobalGaussrate}.

We consider again the OPO as a paradigmatic model.
Focusing on the resonant case $\omega = 0$, the equations can be solved analytically to find the SCGF $C(\lambda) =  \frac{1}{4} \biggl( 2\kappa -  \sqrt{\kappa ^2+4 \kappa  e^{ \I \lambda } \chi  +4 \chi ^2}  - \sqrt{\kappa ^2-4 \kappa  e^{ \I \lambda } \chi +4 \chi ^2} \biggr)$  from which $\mathtt{J} = \frac{2 \kappa \chi^2 }{\kappa^2 - 4 \chi^2}$ and  $\mathtt{D} = \frac{4 \kappa  \chi ^2 \left(\kappa ^4+2 \kappa ^2 \chi ^2+8 \chi ^4\right)}{\left(\kappa ^2-4 \chi ^2\right)^3}$, which coincide exactly with the values obtained in Ref.~\cite{Landi2024} from direct calculations.
The OPO is a paradigmatic quantum optics model; its photodetection statistics has been studied for a long time, and many analytical results are known~\cite{Vyas1989,Padurariu2012,Arndt2021,Hassler2023,Bultron2024}.
Nonetheless, our formalism makes the analytical derivation of the SCGF extremely straightforward.

Moreover, also the long-time QFI that leads to the corresponding TUR can be computed analytically for $\omega = 0$, obtaining $f = 2 \chi^2/ \kappa $.
The comparison between the two sides of Hasegawa's TUR is shown in Fig.~\ref{fig2}.
We see that the lower bound is tight for small $\chi/\kappa$, where the noise dominates, however it misses the fact that noise blows up again for $\chi \to \kappa / 2$, where the system becomes unstable.
We refer to the Supplemental Material~\cite{supp} for more details on these derivations, where we also show a simple analytical example of the finite-time counting statistics of non-Gaussian initial cat states, obtained by propagating with the same equations four Gaussian components.

\emph{Conclusions.}---In this letter we have derived ODEs for the covariance matrix, first moments, and normalization of a Gaussian operator evolving under a Gaussianity-preserving GME, and have shown how to apply the formalism to topical problems: the evaluation of the environmental QFI with TSMEs and replica GMEs, and full counting statistics with tilted MEs.
These examples only illustrate a subset of potential applications; we list a few promising directions below.

The trace of a state evolved under the no-jump part of a Lindblad generator can be used to obtain the waiting-time distribution, recently investigated in fermionic systems~\cite{Coppola2024} and microwave cavities~\cite{Brange2019b,Portugal2023,Kansanen2025}.
Hierarchies of GMEs arise in the study of first-passage times for continuous measurement currents~\cite{Kewming2024}, with applications to the analysis of quantum clocks~\cite{Singh2025}.
Our method is also compatible with time-dependent generators.
In particular, a hierarchy of time-dependent GMEs describes the reduced dynamics of quantum systems driven by quantum pulses of radiation~\cite{Baragiola2012,Baragiola2017,Gross2022}.
Moreover, time-dependent TSMEs (and replica GMEs) are relevant for sensitivity limits in parameter estimation with classical~\cite{Yang2023e,Khan2025,Chinni2025,Yang2026a} and quantum pulses~\cite{Khanahmadi2023,Albarelli2023a}.

Finally, it would be interesting to extend this approach to the fermionic case in full generality, since the moment equations for Gaussian fermionic dynamics are closely analogous to the bosonic ones~\cite{Purkayastha2022,Barthel2022a}.

\emph{Acknowledgments.}---FA thanks Simone Felicetti for useful discussions.
The authors acknowledge useful discussions with Lorenzo Braccini and Alessio Serafini, and especially for pointing out the close connection to their work on superpositions of Gaussian processes~\cite{Braccini2025a}.
AI tools (ChatGPT and Codex) were used to assist with checking analytical calculations, implementing numerical code for the applications presented in this work, preparing the code for public release on GitHub, and improving the language of the manuscript.
The authors reviewed all AI-assisted output and take full responsibility for the content of the manuscript and the accompanying code.

\emph{Data availability.}---The Julia implementation of the Gaussian moment equations, together with the scripts and data for reproducing the figures, is publicly available in Ref.~\cite{GaussianGMEsCode}.

\bibliography{2026GeneralizedMEGaussian}

\onecolumngrid
\begin{center}
    \large\bfseries End Matter
\end{center}
\twocolumngrid

\emph{Sketch of the moment equations derivation.}---We consider the GME in Eqs.~\eqref{eq:GMEmu},~\eqref{eq:quadratic-A-C}
and~\eqref{eq:LmuR_linear}, for a finite number of modes and an initial Gaussian operator whose characteristic function has finite parameters and $\operatorname{Re}\sigma_0\succ0$.
For any such Gaussian operator $\hat{\mu}$, the characteristic function of the right-hand side of Eq.~\eqref{eq:GMEmu} is a polynomial of degree at most two in the phase-space variables multiplying $\chi_{\hat{\mu}}$.
Any such function is a tangent vector to the unnormalized Gaussian family with complex parameters, cf.~\cite{Corney2003a}.
Matching the constant, linear, and quadratic coefficients
with those of the time derivative of the Gaussian
parametrization therefore yields
Eqs.~\eqref{eq:dotsigma}--\eqref{eq:doteta}.
Their integration reconstructs a trace-class Gaussian
solution of the GME on every interval where the parameters
remain finite and $\operatorname{Re}\sigma\succ0$.

To derive these equations, it is convenient to start from the action of some linear and quadratic superoperators on the characteristic function:
\begin{equation}
    \chi_{\left[ \hat{r}_j , \hat{o} \right]} (\tilde{r}) 
     =  \Omega_{jk} \tilde{r}_k \chi_{\hat{o}} (\tilde{r})  , \; \; \chi_{\left\{ \hat{r}_j , \hat{o} \right\}} (\tilde{r}) 
    =  - 2 \I \partial_{\tilde{r}_j} \chi_{\hat{o}} (\tilde{r})  
    \label{eq:comm_anti_comm_chi_rj}
\end{equation}
\begin{equation}
    \label{eq:comm_anti_comm_chi_rkrj_old}
    \begin{aligned}
    \chi_{\left[ \hat{r}_k \hat{r}_j , \hat{o} \right]}
     = & - \I\left(  \Omega_{j j'} \tilde{r}_{j'} \partial_{\tilde{r}_k} +  \Omega_{k k'} \tilde{r}_{k'}\partial_{\tilde{r}_j}  \right) \chi_{\hat{o}}  \\
     \chi_{\left\{ \hat{r}_k \hat{r}_j , \hat{o} \right\}}
    = & - 2 \partial_{\tilde{r}_k} \partial_{\tilde{r}_j} \chi_{\hat{o}} + \I\Omega_{kj}\chi_{\hat{o}} + \frac{1}{2}  \Omega_{kk'} \Omega_{jj'} \tilde{r}_{k'} \tilde{r}_{j'}\chi_{\hat{o}} .
    \end{aligned}
\end{equation}

\begin{align}
   & \chi_{\hat{r}_{j} \hat{o} \hat{r}_{k}} (\tilde{r}) =  \left[ -\I\partial _{\tilde{r}_{k}}-%
    \frac{\Omega _{kk'}}{2}\tilde{r}_{k'}\right]
    \left[ - i \partial_{\tilde{r}_j} +  \frac{\Omega_{jj'}}{2} \tilde{r}_{j'} \right] \chi_{\hat{o}}(\tilde{r}) \\
    & = \biggl[ - \partial_{\tilde{r}_k} \partial_{\tilde{r}_j} -  \frac{1}{4} \Omega _{kk'}\Omega_{jj'} \tilde{r}_{k'} \tilde{r}_{j'}
    \\
    & \qquad + \I \left( \frac12 \Omega_{k k'} \tilde{r}_{k'} \partial_{\tilde{r}_j}  - \frac12 \Omega_{j j'} \tilde{r}_{j'} \partial_{\tilde{r}_k} - \frac12 \Omega_{jk} \right) \biggr] \chi_{\hat{o}}(\tilde{r}) \nonumber \, .
\end{align}
These identities can be applied to map Eq.~\eqref{eq:GMEmu} with the quadratic and linear operators in Eqs.~\eqref{eq:quadratic-A-C} and~\eqref{eq:LmuR_linear} onto a partial differential equation for the characteristic function of $\hat{\mu} = e^{-\xi} \hat{\nu}$.
By remembering that
\begin{equation}
    \frac{d \hat{\mu}}{dt} = -\frac{d \xi }{dt} e^{- \xi } \hat{\nu} + e^{- \xi  } \frac{d \hat{\nu} }{d t} ,
    \label{eq:dmu}
\end{equation}
and the characteristic function of $\hat{\nu}$ in Eq.~\eqref{eq:nu_Gauss_charfun}, one has to only consider the action of partial derivatives onto a Gaussian characteristic function $\chi_G$:
\begin{align}
& \frac{\partial_{\tilde{r}_j} \chi_G}{\chi_G} =  \I d_j - \frac{1}{2} \sigma_{jj'} \tilde{r}_{j'} , \\
& \frac{\partial_{\tilde{r}_k} \partial_{\tilde{r}_j} \chi_G}{\chi_G} =  \left( \I d_k - \frac{1}{2} \sigma_{kk'} \tilde{r}_{k'} \right) \left( \I d_j - \frac{1}{2} \sigma_{jj'} \tilde{r}_{j'} \right) - \frac{1}{2} \sigma_{jk}  \nonumber \\ 
& =  - d_k d_j + \frac{1}{4} \sigma_{kk'}\tilde{r}_{k'} \sigma_{jj'}\tilde{r}_{j'} \\ 
& \quad -\frac{\I}{2} \sigma_{kk'} \tilde{r}_{k'} d_j -\frac{\I}{2} d_k \sigma_{jj'} \tilde{r}_{j'}  -\frac{1}{2} \sigma_{jk} , \nonumber \\
& \frac{\dot{\chi}_G}{\chi_G} =  \I \tilde{r}_p \dot{d}_p - \frac{1}{4} \dot{\sigma}_{lm} \tilde{r}_l \tilde{r}_m  =   \I \tilde{\vec{r}}^\tran \dot{\vec{d}} - \frac{1}{4} \tilde{\vec{r}}^\tran \dot\sigma \tilde{\vec{r}} \, .
\end{align}
The final ODEs in Eqs.~\eqref{eq:dotsigma},~\eqref{eq:dotd} and \eqref{eq:doteta} can be obtained by combining all the equations above and laboriously matching all the coefficients of the different powers of the phase-space variables $\tilde{r}_j$.
In doing so, it is useful to notice that when expressions such as $\vec{\tilde{r}}^\tran M \vec{\tilde{r}}$ appear, these are actually constraining only the symmetric part $\operatorname{Sym}[M]$.
 
\emph{Generalized replica master equation.}---We consider a monitored jump operator $\hat{J}$ and unmonitored jump operators $\hat{L}_k$.
For this scenario, Yang et al.~\cite{Yang2026a} derived a generalized replica master equation
\begin{equation}
    \frac{d \hat{\mu} }{dt} = \sum_{\alpha=1}^{m+2} \mathcal{L}^{(\alpha)}_{\theta_{\alpha}} \hat{\mu} + \sum_{\alpha=1}^{m+2} \hat{J}^{(\alpha+1)}_{\theta_{\alpha+1}} \hat{\mu} \hat{J}^{(\alpha)\dag}_{\theta_{\alpha}}
\end{equation}
where 
\begin{align}
    \mathcal{L}_{\theta} \hat{\mu} = -\I [ \hat{H}_{\theta} , \hat{\mu} ]  - \frac{1}{2} \left\{ \hat{J}^\dag_{\theta} \hat{J}_{\theta} , \hat{\mu} \right\} 
     + \sum_k \mathcal{D}[\hat{L}_{k,\theta}] \hat{\mu} \, .
\end{align}
is the no-jump evolution for the monitored jump operator, while $\mathcal{D}[\hat{L}_{k,\theta}] \hat{\mu} = \hat{L}_{k,\theta} \hat{\mu} \hat{L}_{k,\theta}^\dag - \{ \hat{L}_{k,\theta}^\dag \hat{L}_{k,\theta}  , \hat{\mu} \} / 2 $ are the standard trace-preserving Lindblad dissipators. 
The replicas are connected by a periodic boundary condition $m+3 \equiv 1$.
This GME for replicas yields the Bargmann invariants 
\begin{equation}
\label{eq:bargmann_invariant}
B(\Theta)\; := \;\Tr\!\Big[\,\Xi(\theta_1) \,\Xi(\theta_2) \cdots \Xi(\theta_{m+2})\,\Big]  = \Tr[ \hat{\mu} ] ,  
\end{equation}
where $\Theta=(\theta_1,\ldots,\theta_{m+2})$.
Here $\Xi(\theta)$ is the reduced state of the continuum of modes coupled to the system via $\hat{J}_{\theta}$ (dubbed \emph{the waveguide} in Ref.~\cite{Yang2026a}), obtained after tracing out both the system and the inaccessible environments coupled via the other jump operators.

A crucial observation is that the QFI of the waveguide can be expressed in terms of Bargmann invariants, via a convergent series of approximants:
\begin{equation}
\label{eq:In_series}
\mathcal{Q}_{N,\theta}
=\sum_{m=0}^{N}(-1)^m \binom{N+1}{m+1}\,
\frac{f_m(\theta)}{\Lambda(\theta)^{m+1}},
\end{equation}
where
\begin{equation}
\label{eq:fm_def}
f_m(\theta)
=\sum_{l=0}^{m} D_l^{\,m}\;
\partial_{\alpha}\partial_{\beta}\,
\Tr\!\Big[\Xi(\alpha)^{\,l+1}\,\Xi(\beta)^{\,m-l+1}\Big]\Big|_{\alpha=\beta=\theta},
\end{equation}
and
\begin{equation}
\label{eq:Lambda_def}
\Lambda(\theta)\;:=\;2\,\Big(\tr[\Xi(\theta)^2]\Big)^{1/2}.
\end{equation}
The coefficients $D_l^{\,m}$ are
\begin{align}
\label{eq:Dlm_def}
D_l^{\,m}
=2\binom{m}{l}-\binom{m}{l+1}-\binom{m}{l-1},
\end{align}
where $\binom{m}{l}=0\;\;\text{for}\;\; l<0\ \text{or}\ l>m$.
While $\mathcal{Q}_{N,\theta}$ converges to the QFI of the waveguide as $N\to \infty$, for each $N$ this quantity is a valid lower bound.
Thus, even when convergence is slow, evaluating $\mathcal{Q}_{N,\theta}$ may still provide useful information on the estimation task, e.g. showing suboptimality of specific measurements.

Let us consider a quadratic $\hat{H}_{\theta}$, a linear monitored jump operator $\hat{J}_{\theta} = \mathbb{j}(\theta) \hat{\vec{r}} = \sum_{k=1}^{2 n} \mathbb{j}_k(\theta)  \hat{r}_k$, and linear unmonitored jump operators $\hat{L}_{k,\theta} = \sum_{j=1}^{2 n} \mathbb{L}_{kj}(\theta)  \hat{r}_j$.
Notice that $\mathbb{j}(\theta)$ is treated as a row vector, i.e. a $1{\times}2n$  matrix.
Each replica is an $n$-mode bosonic system, and the system of $m + 2$ replicas has $(m + 2)n$ modes.
The coefficient matrices related to commutator and anticommutator correspond to the independent physical evolution of each replica, leading to block-diagonal matrices and vectors:
\begin{align}
    \mathbb{C} &= -\I  \bigoplus_{\alpha=1}^{m+2} \mathbb{H}(\theta_\alpha)   \qquad \mathbb{c}  = -\I  \bigoplus_{\alpha=1}^{m+2} \mathbb{h}(\theta_\alpha) \\
    \mathbb{A} &= - \bigoplus_{\alpha=1}^{m+2} \left( \mathbb{j}^\dag(\theta_\alpha) \mathbb{j}(\theta_\alpha) + \mathbb{L}^\dag(\theta_\alpha)\mathbb{L}(\theta_\alpha) \right) \qquad \mathbb{a} & = 0 
\end{align}
The nontrivial part of the generalized replica master equation resides only in the ``jump'' term. 
The $\mathbb{K}$ matrix diagonal blocks correspond to the physical evolution of each replica with the unmonitored jump operators:
\begin{align}
    \mathbb{K}_{\alpha,\alpha} = \left( \mathbb{L}^\dag(\theta_\alpha)\mathbb{L}(\theta_\alpha) \right)^\tran  \, ,
\end{align}
while the off-diagonal blocks are related to the monitored jump operators, and they are nonzero only for nearest-neighbour replicas:
\begin{equation}
\mathbb{K}_{\alpha+1,\alpha} = \left( \mathbb{j}^\dag (\theta_{\alpha})  \mathbb{j}(\theta_{\alpha+1})\right)^\tran  \, .
\end{equation}
From these expressions, the matrices in Eq.~\eqref{eq:AandDdef} are easily obtained.

\begin{figure}
\includegraphics{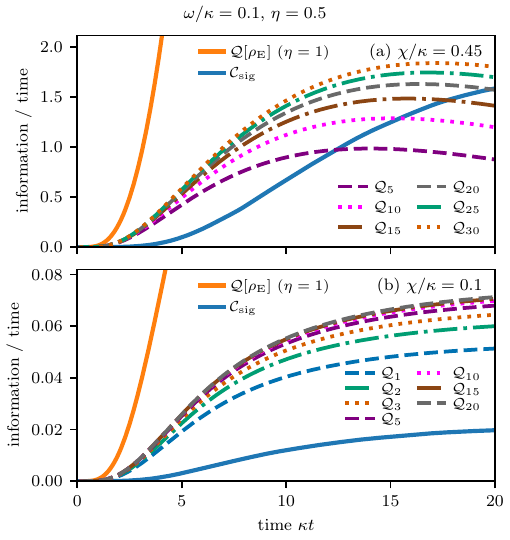}
\caption{Fisher information rates for $\omega$ estimation with an inefficiently ($\eta = 0.5$) monitored optical parametric oscillator (OPO), as a function of time;  
$\chi/\kappa = 0.45$ in the top panel, and $\chi/\kappa = 0.1$ in the bottom panel.}
\label{fig3}
\end{figure}

As an application, we focus again on frequency estimation with a continuously monitored OPO, this time with inefficient detection, which amounts to having a monitored jump operator $\hat{J} = \sqrt{\kappa \eta} \hat{a}$ and an unmonitored $\hat{L} = \sqrt{\kappa(1-\eta)} \hat{a} $.
In Fig.~\ref{fig3} we show various approximations for the QFI of the detectable environment, i.e. $\mathcal{Q}_{N}$ for different values of $N$, together with the environment TSME-QFI (which corresponds to perfect efficiency $\eta =1$), and the homodyne signal CFI $\mathcal{C}_{\mathrm{sig}}$.
First, we see that the TSME is a very loose upper bound, as expected from more general noisy quantum metrology bounds~\cite{Gorecki2025a,Blinova2026}.
Second, we see that more and more terms are needed for convergence of $\mathcal{Q}_{N}$ to the actual QFI, both as time increases and as the stability threshold $\chi/\kappa = 1/2$ is approached.
We evaluate the partial derivatives in Eq.~\eqref{eq:fm_def} by integrating sensitivity equations obtained by differentiating the replica moment equations with respect to the Hamiltonian parameter.

\emph{Full counting statistics for diffusive measurements.}---For diffusive measurements we have~\cite{Landi2024}
\begin{align}
    \dot{\hat{\mu}} =&  -\I [ \hat{H} , \hat{\mu} ] + \sum_k \left( \hat{J}_k \hat{\mu} \hat{J}_k^\dag - \frac{1}{2} \left\{ \hat{J}_k^\dag \hat{J}_k , \hat{\mu} \right\} \right)  \\
    &  + \I \lambda  \sum_k w_k \left( e^{- \I \phi_k} \hat{J}_k \hat{\mu} + e^{+ \I \phi_k } \hat{\mu} {\hat{J}_k^\dag} \right) - \frac{\lambda^2}{2} \left( \sum_k w_k^2 \right) \hat{\mu} . \nonumber
\end{align}
Introducing 
\begin{equation}
\hat{F} \;:=\; \sum_k w_k e^{-\I\phi_k}\, \hat{J}_k,
\; \;
\hat{G} \;:=\; \sum_k w_k e^{+\I\phi_k}\, \hat{J}_k^\dag, 
\end{equation}
so that
\begin{eqnarray}
    \hat{C} = - \I \hat{H} + \frac{ \I \lambda}{2} \left( \hat{F} - \hat{G} \right) 
\end{eqnarray}
and
\begin{eqnarray}
    \hat{A} = -\frac{1}{2} \sum_k \hat{J}_k^\dag \hat{J}_k + \frac{\I \lambda}{2} \left( \hat{F} + \hat{G} \right) - \frac{\lambda^2}{4} \left( \sum_k w_k^2 \right) \id \, ,
\end{eqnarray}
since $[\,\frac{\I\lambda}{2}(\hat{F}-\hat{G}),\hat{\mu}]+\{\,\frac{\I \lambda}{2}(\hat{F}+\hat{G}),{\hat{\mu}}\}= \I \lambda(\hat{F}\hat{\mu}+\hat{\mu} \hat{G})$ and the scalar term $-\frac{\lambda^2}{2}\left(\sum_k w_k^2\right)\hat{\mu}$ is reproduced by adding $-\frac{\lambda^2}{4}\left(\sum_k w_k^2\right)\id$ inside $\hat A$, since $\{\id,\hat{\mu}\}=2{\hat{\mu}}$.

Apart from the scalar term, only the linear terms are modified.
Introducing
\begin{equation}
    \mathbb{f} = \sum_k w_k e^{-\I \phi_k} {\vec{\mathbb{j}}_k}  \qquad \mathbb{g} = \sum_k w_k e^{+\I \phi_k} {\vec{\mathbb{j}}_k^\ast} ,
\end{equation}
where $\hat{J}_k = \vec{\mathbb{j}}_k^\tran \hat{\vec{r}}$, i.e. ${\vec{\mathbb{j}}_k^\tran}$ are the rows of ${\mathbb{J}}$, we have
\begin{equation}
    \mathbb{c} =  -\I \mathbb{h} + \frac{ \I \lambda}{2}\, {\Omega}(\mathbb f-\mathbb g),\qquad
\mathbb{a} = \frac{\I \lambda}{2}\,{\Omega}(\mathbb f+\mathbb g) \, .
\end{equation}
Finally, the term proportional to the identity can be included either by adding the purely imaginary antisymmetric matrix $\frac{\I\lambda^2}{2n}\left(\sum_k w_k^2\right)\Omega$ to $\mathbb{A}$, or by adding $\frac{\lambda^2}{2}\sum_k w_k^2$ directly to the equation for $\dot{\xi}$.

\onecolumngrid 
\clearpage
\appendix
\widetext

\centerline{\large \textbf{Supplemental material}}
\section{Deterministic equations for the unravelling QFI of Gaussian systems}
\label{app:deterministic_unravelling}
The signal CFI for Gaussian systems can be evaluated by solving stochastic differential equations for the first and second moments and their parametric derivatives~\cite{Genoni2017}.
While this phase-space formulation already permits an effective evaluation by sampling quantum trajectories numerically, a deterministic moment formulation was recently derived in Ref.~\cite[Supplemental Material, Sec.~IV.A]{Zhang2025Backaction}.
Here, we report the deterministic moment equations for general linear Gaussian Markov dynamics and show that the same moments also give the average conditional-state QFI, and thus enable the evaluation of the unravelling QFI, $\mathcal{Q}_{\mathrm{unr},\theta} := \mathcal{C}_{\mathrm{sig},\theta} + \mathbbm{E}_{\mathrm{traj}}\left[ \mathcal{Q}[\rho_{\mathrm{S},\theta}^{\mathrm{(cond)}}]  \right]$, by solving ODEs, rather than stochastic equations.

We consider an arbitrary $n$-mode Gaussian initial state, quadratic Hamiltonians (including linear drives), linear couplings to Markovian baths, and general-dyne monitoring~\cite{Genoni2017}.
The coefficients and measurement settings may depend on time, but do not depend on the observed current.
We assume that the measurement-noise covariance matrix is independent of the estimated parameter $\theta$, so that a parameter-independent normalization gives the following stochastic evolution for the first moments $\vec{d}$ and measurement record $\vec{y}$:
\begin{align}
    d\vec{d} &= (A\vec{d}+\vec{a})\,dt+K_\sigma\,d\vec{W},
    & d\vec{y} &= (H\vec{d}+\vec{h})\,dt+d\vec{W}, \nonumber\\
    \dot\sigma &= A\sigma+\sigma A^\tran+D-2K_\sigma K_\sigma^\tran,
    & \mathbbm{E}[d\vec{W}\,d\vec{W}^\tran] &= \id\,dt.
    \label{eq:unr_filter}
\end{align}
Here $\sigma$ is the conditional system covariance matrix, and for compactness we have introduced the matrix $K_\sigma=(N-\sigma B)/\sqrt{2}$; 
$B$ and $N$ are deterministic matrices specified by the system--bath coupling and the measurement scheme~\cite{Genoni2017,Serafini2023}; they are also related to the matrix $H=-\sqrt{2}B^\tran$ appearing in the equations.

Parametric differentiation at a \emph{fixed measurement record} gives
\begin{align}
    \partial_\theta d\vec{W} &=-\bigl[(\partial_\theta H)\vec{d}+H\partial_\theta\vec{d}+\partial_\theta\vec{h}\bigr]\,dt, \nonumber\\
    \frac{d}{dt}(\partial_\theta\sigma) &=(\partial_\theta A)\sigma+\sigma(\partial_\theta A)^\tran
    +A(\partial_\theta\sigma)+(\partial_\theta\sigma)A^\tran+\partial_\theta D -2\bigl[(\partial_\theta K_\sigma)K_\sigma^\tran+K_\sigma(\partial_\theta K_\sigma)^\tran\bigr].
    \label{eq:unr_sensitivity}
\end{align}
The derivative of $K_\sigma$ includes its dependence on the conditional covariance matrix $
    \partial_\theta K_\sigma=\frac{\partial_\theta N-(\partial_\theta\sigma)B-\sigma(\partial_\theta B)}{\sqrt{2}}$.
Both $\sigma$ and $\partial_\theta\sigma$ are deterministic.
Introducing an additional constant component to the equations includes arbitrary displacements and drives:
\begin{align}
    d\vec{z} &= F\vec{z}\,dt+G\,d\vec{W},
    \qquad \vec{z}=\begin{pmatrix}\vec{d}\\\partial_\theta\vec{d}\\1\end{pmatrix}, \nonumber\\
    F &=\begin{pmatrix}A&0&\vec{a}\\\partial_\theta A-K_\sigma\partial_\theta H&A-K_\sigma H&\partial_\theta\vec{a}-K_\sigma\partial_\theta\vec{h}\\0&0&0\end{pmatrix},
    \qquad G=\begin{pmatrix}K_\sigma\\\partial_\theta K_\sigma\\0\end{pmatrix}.
    \label{eq:unr_augmented_filter}
\end{align}
Since $F$ and $G$ are deterministic, It\^o's rule yields a closed deterministic differential equation for the ensemble second moment $V=\mathbbm{E}_{\mathrm{traj}}[\vec{z}\vec{z}^\tran]$:
\begin{equation}
    \dot V=FV+VF^\tran+GG^\tran,
    \qquad V(0)=\vec{z}_0\vec{z}_0^\tran,
    \quad \vec{z}_0=(\vec{d}_0^\tran,(\partial_\theta\vec{d}_0)^\tran,1)^\tran.
    \label{eq:unr_moment_closure}
\end{equation}
The covariance matrix initial conditions are $\sigma(0)=\sigma_0$ and $(\partial_\theta\sigma)(0)=\partial_\theta\sigma_0$; neither thermal nor parameter-independent initial states are required.
Since the current $\vec{y}$ has a conditional mean part and a Gaussian white noise part, the signal CFI $\mathcal{C}_{\mathrm{sig}}$, being the expected value of the score function squared, is obtained by integrating the corresponding information rate over time.
Because the noise variance is constant, equal to $dt$, the rate at which the signal CFI accumulates over time (i.e. the signal CFI rate) reduces purely to the expected squared norm of the parameter derivative of the conditional mean signal.
This derivative is $\partial_\theta(H\vec{d}+\vec{h})=J\vec{z}$, where we have introduced the block matrix $J=(\partial_\theta H \;\; H \;\; \partial_\theta\vec{h} )$.
In formula: 
\begin{align}
    \dot{\mathcal{C}}_{\mathrm{sig},\theta} &= \mathbbm{E}_{\mathrm{traj}}[(J \vec{z})^T(J \vec{z})] = \tr(JVJ^\tran),
    \qquad \mathcal{C}_{\mathrm{sig},\theta}(0)=0
     \label{eq:signalCFI}
\end{align}
Let us now consider the average of the conditional states QFI  $\mathbbm{E}_{\mathrm{traj}}\left[ \mathcal{Q}[\rho_{\mathrm{S},\theta}^{\mathrm{(cond)}}]  \right]$: we remind ourselves that the QFI of multimode Gaussian states~\cite{Safranek2019,Chang2026} is the sum of a term depending on the covariance matrix only $\mathcal{Q}_{\mathrm{cov}}$, plus a term depending also on the first moments 
\begin{equation}
    \mathcal{Q}_{\mathrm{fm}}= 2 (\partial_\theta\vec{d})^\tran \sigma^{-1} (\partial_\theta\vec{d}) \,.
\end{equation}
The evolution of the covariance matrix is deterministic, thus we only need to average $\mathcal{Q}_{\mathrm{fm}}$ over the trajectories, and we eventually obtain
\begin{align}
  \mathbbm{E}_{\mathrm{traj}}\left[ \mathcal{Q}[\rho_{\mathrm{S},\theta}^{\mathrm{(cond)}}]  \right] = \mathcal{Q}_{\mathrm{cov}}  + 2\, \mathbbm{E}_{\mathrm{traj}}\left[ {(\partial_\theta\vec{d})}^\tran \sigma^{-1} {(\partial_\theta\vec{d})} \right] 
  = \mathcal{Q}_{\mathrm{cov}}  + 2\tr(\sigma^{-1}P) 
  \, , 
\end{align}
where $P=\mathbbm{E}_{\mathrm{traj}}[(\partial_\theta\vec{d})(\partial_\theta\vec{d})^\tran]$ is a submatrix of $V$ and thus evolves \emph{deterministically} according to Eq.~\eqref{eq:unr_moment_closure}.
When all the symplectic eigenvalues of the conditional covariance matrix exceed one, i.e. the conditional Gaussian state has full symplectic rank, the covariance-dependent term of the conditional states QFI reads~\cite{Safranek2019,Chang2026}
\begin{equation}
    \mathcal{Q}_{\mathrm{cov}}=\frac12\operatorname{vec}({\partial_\theta\sigma})^\tran
    (\sigma\otimes\sigma-\Omega\otimes\Omega)^{-1}\operatorname{vec}({\partial_\theta\sigma}),
    \label{eq:unr_covariance_qfi}
\end{equation}
where $\operatorname{vec}$ stacks matrix columns into a single vector.
For a pure-state family, $\mathcal{Q}_{\mathrm{cov}}=\tr[(\sigma^{-1}{\partial_\theta\sigma})^2]/4$; for quantum states with some pure modes, i.e. some symplectic eigenvalues equal to one, one can obtain the QFI by taking a limit~\cite{Safranek2019} or by explicit formulas in terms of a symplectic diagonalisation of the covariance matrix~\cite{Serafini2023}.

To conclude, the unravelling QFI can be efficiently evaluated via the formulas
\begin{align}
    \dot{\mathcal{C}}_{\mathrm{sig},\theta} &= \tr(JVJ^\tran),
    \qquad \mathcal{C}_{\mathrm{sig},\theta}(0)=0, \nonumber\\
    \mathcal{Q}_{\mathrm{unr},\theta} &= \mathcal{C}_{\mathrm{sig},\theta}
    +\mathcal{Q}_{\mathrm{cov}}+2\tr\left[\sigma^{-1} P \right].
    \label{eq:unr_deterministic_information}
\end{align}

For the OPO example we consider in the manuscript and that we will detail in a following section, we have $\theta=\omega$, $\vec{a}=\vec{h}=0$, $N=B$, $K_\sigma=(B-\sigma B)/\sqrt{2}$, and $H=-\sqrt{2}B^\tran$, with $B=-\sqrt{\eta\kappa}\,\vec{v}_\phi\vec{v}_\phi^\tran$ for homodyne detection, $\vec{v}_\phi=(\cos\phi,-\sin\phi)^\tran$, and $B=-\sqrt{\eta\kappa/2}\,\id_2$ for heterodyne detection.
Eq.~\eqref{eq:signalCFI} then reduces to $\dot{\mathcal{C}}_{\mathrm{sig},\omega}=2\tr\left[B^\tran P B\right]$.

Finally, we remark that this argument generally does not apply when record-dependent measurement settings are considered.

\section{Trace norm of a Gaussian operator}

We start the derivation by separating the $\xi$ contribution:
\begin{equation}
\label{eq:tracenorm_mu_supp}
\Vert \hat{\mu} \Vert_1 = \Tr\left[ \sqrt{\hat{\mu}^\dag \hat{\mu}}  \right] = e^{- \Re[\xi]} \Tr\left[ \sqrt{\hat{\nu}^\dag \hat{\nu}}  \right]  .
\end{equation}
Then, we introduce the normalized state $\hat{\rho} = \hat{\nu}^\dag \hat{\nu} / p $ with $p := \Tr[ \hat{\nu}^\dag \hat{\nu} ]$, from which 
\begin{equation}
\label{eq:tracenorm_nu_supp}
    \left\Vert \hat{\nu} \right\Vert_1 = \sqrt{ p } \Tr[ \hat{\rho}^{1/2} ].
\end{equation}
Crucially, the product of Gaussian operators remains Gaussian, as will be evident from the explicit construction in the next subsection.
Therefore $\hat{\rho}$ is a proper Gaussian state and a formula for the trace of the square root of an $n$-mode Gaussian state is known in terms of its symplectic eigenvalues $\nu_k$~\cite{Seshadreesan2018}:
\begin{equation} 
\Tr\left[ \hat{\rho}^{1/2} \right] = 2^{-n/2}\prod_{k=1}^n\Big(\sqrt{\nu_k+1}+\sqrt{\nu_k-1}\Big) \, .
\end{equation}
Thus, the remaining ingredients are expressions for the normalization $p$ and covariance matrix of $\hat{\rho}$.
The explicit derivation of these results is in the following subsections.

\subsubsection{Gaussian-integral evaluation of the moments of $K=\hat\nu^\dagger \hat\nu$}
\label{app:nu_dag_nu_gauss_integrals}

For a trace-class operator $\hat O$ we use the symmetrically-ordered (Weyl) characteristic function:
\begin{equation}
\chi_{\hat O}(\bs):=\Tr\!\left[ \hat{O} \,\hat{D}(-\bs)\right],
\qquad
\hat{D}(\bs):=\exp\!\big(\I\,\bs^{\mathsf T}\Omega\rvec\big),
\qquad \bs\in \mathbb R^{2n}.
\end{equation}
We consider a (generally non-Hermitian) Gaussian operator $\hat\nu$, as in the main text, specified by
\begin{equation}
\chi_{\hat{\nu}}(\bs)
=
\exp\!\left[-\frac14\,\bs^{\mathsf T}\Omega^{\mathsf T}\sigma\,\Omega\,\bs\right]\,
\exp\!\left[ \I \,\bs^{\mathsf T}\Omega^{\mathsf T}\vec{d}\right],
\label{eq:chi_nu_app}
\end{equation}
where $\sigma=\sigma^{\mathsf T}\in\mathbb C^{2n\times 2n}$ and $\vec{d}\in\mathbb C^{2n}$.
We assume $\Tr[ \hat{\nu}] = \chi_{\hat\nu}(0)=1$ and that all Gaussian integrals converge, in particular
\begin{equation}
\Re[\sigma] \succ 0.
\end{equation}
In the following, we denote $ \sigma_R := \Re[\sigma]$ and $\sigma_I:= \Im[\sigma]$ (both real symmetric), and decompose
$\vec{d}=\vec{d}_R + \I \,\vec{d}_I$ with $\vec{d}_{R,I}\in\mathbb R^{2n}$.

First, we recall the Weyl relation among displacement operators~\cite{Serafini2023}:
\begin{equation}
\hat{D}(\bs) \hat{D}(\yb)=\exp\!\left(-\frac{\I}{2}\bs^{\mathsf T}\Omega\yb\right)\,\hat{D}(\bs+\yb),
\label{eq:weyl_relation_app}
\end{equation}
then we recall the definition of the inverse Weyl transform~\cite{Serafini2023}
\begin{equation}
\hat{O}=\frac{1}{(2\pi)^n}\int_{\mathbb R^{2n}} d^{2n}\bs\;\chi_{\hat O}(\bs)\, \hat D(\bs) \, ,
\label{eq:inverse_weyl_app}
\end{equation}
which allows to write the operator in terms of the characteristic function.
Moreover, for any operator $\hat{O}$ one has
\begin{equation}
\chi_{\hat{O}^\dagger}(\bs)=\chi_{\hat{O}}(-\bs)^*,
\label{eq:chi_adjoint_app}
\end{equation}
which follows directly from $\hat{D}(\bs)^\dagger=\hat{D}(-\bs)$ and cyclicity of the trace (which holds since $\hat O$ is trace-class and $\hat D(\bs)$ is bounded).

Our goal is to compute the Gaussian parameters of
\begin{equation}
\hat{K} := \hat{\nu}^\dagger \hat{\nu},
\end{equation}
namely its trace $p=\Tr[\hat{K}]$, its covariance matrix $\sigma_{\hat{K}}$, and its first-moment vector $\vec{d}_{\hat{K}}$ appearing in the Gaussian form of $\chi_{\hat{K}}$.
To achieve this goal, we start by recalling that for trace-class operators $\hat{A},\hat{B}$,
\begin{equation}
\chi_{\hat{A}\hat{B}}(\zb)=\frac{1}{(2\pi)^n}\int_{\mathbb R^{2n}} d^{2n}\bs\;
\chi_{\hat A}(\bs)\,\chi_{\hat B}(\zb-\bs)\;
\exp\!\left(-\frac{\I}{2}\,\bs^{\mathsf T}\Omega\,\zb\right).
\label{eq:twisted_convolution_app}
\end{equation}
This statement comes from inserting \eqref{eq:inverse_weyl_app} for $\hat{A}$ and $\hat{B}$, then applying the relation in Eq.~\eqref{eq:weyl_relation_app},
and finally taking the trace with a displacement operator, i.e. $\Tr[\cdot\,\hat{D}(-\zb)]$.
The trace enforces a delta constraint $\bs+\yb=\zb$, leaving the phase factor in \eqref{eq:twisted_convolution_app}.

Using \eqref{eq:chi_adjoint_app}, the characteristic function of $\hat{\nu}^\dagger$ is
\begin{equation}
\chi_{\hat{\nu}^\dagger}(\bs)=\chi_{\hat{\nu}}(-\bs)^*
=
\exp\!\left[-\frac14\,\bs^{\mathsf T}\Omega^{\mathsf T}\sigma^*\,\Omega\,\bs\right]\,
\exp\!\left[\I\,\bs^{\mathsf T}\Omega^{\mathsf T}\vec{d}^{\,*}\right].
\label{eq:chi_nu_dag_app}
\end{equation}
Applying \eqref{eq:twisted_convolution_app} with $\hat{A}=\hat{\nu}^\dagger$ and $\hat{B}=\hat{\nu}$ yields
\begin{equation}
\chi_{\hat K}(\zb)=\frac{1}{(2\pi)^n}\int d^{2n}\bs\;
\chi_{\hat{\nu}^\dagger}(\bs)\,\chi_{\hat{\nu}}(\zb-\bs)\,
\exp\!\left(-\frac{\I}{2}\,\bs^{\mathsf T}\Omega\,\zb\right).
\label{eq:chiK_integral_app}
\end{equation}
Since the integrand is Gaussian in $\bs$, the result is Gaussian in $\zb$ and can be put in the form
\begin{equation}
\chi_{\hat{K}}(\zb)=p\,
\exp\!\left[-\frac14\,\zb^{\mathsf T}\Omega^{\mathsf T}\sigma_{\hat{K}}\,\Omega\,\zb\right]\,
\exp\!\left[\I\,\zb^{\mathsf T}\Omega^{\mathsf T}\vec{d}_{\hat{K}}\right].
\label{eq:chiK_form_app}
\end{equation}
Dividing by the constant $p=\chi_{\hat{K}}(0)=\Tr[\hat{K}]$ produces the characteristic function of the normalized state
$\hat{\rho}:=\hat{K}/\Tr[\hat{K}]$; importantly, the Gaussian parameters satisfy
\begin{equation}
\sigma_{\hat \rho} = \sigma_{\hat K} ,\qquad \vec{d}_{\hat \rho} = \vec{d}_{\hat K},
\label{eq:rho_params_same_app}
\end{equation}
since normalization only rescales $\chi$ by a $\zb$-independent factor.

Notice that the quantity $\mathrm{Tr}(\hat{\rho}^{1/2})$ depends only on the symplectic spectrum of $\sigma_{\hat\rho}$ and is independent of first moments (displacements).
Thus, the first moments of $\hat{\nu}$ enter the trace norm only through the normalization $p$.

\subsubsection{Closed formulas}

\paragraph{(i) Normalization.}
Using the expressions for the trace in terms of phase-space integrals~\cite{Serafini2023} we have $\Tr[ \hat{\nu}^\dagger \hat{\nu} ]=\frac{1}{(2\pi)^n}\int d^{2n}\bs\,|\chi_{\hat{\nu}}(\bs)|^2$. By using the explicit expression of $\chi_{\hat{\nu}}$ in Eq.~\eqref{eq:chi_nu_app} and solving the Gaussian integral, one obtains
\begin{equation}
p = \Tr[\hat{\nu}^\dagger \hat{\nu}] 
=
\frac{1}{\sqrt{\det(\sigma_R)}}\;
\exp\!\left[2\,\vec{d}_I^{\mathsf T}\sigma_R^{-1}\vec{d}_I\right].
\label{eq:p_supp}
\end{equation}
Notice that only $\Re[\sigma]$ and $\Im[\vec{d}]$ enter this quantity.

\paragraph{(ii) Covariance matrix.}
Evaluating the Gaussian integral \eqref{eq:chiK_integral_app} by completing the square (using
$\Re\sigma\succ0$ to ensure convergence) gives the following explicit expression for the covariance matrix:
\begin{equation}
\sigma_{\hat{K}}
=
\sigma-\frac12\,(\sigma +\I \Omega)\,\sigma_R^{-1}\,(\sigma- \I \Omega).
\label{eq:sigmaK_complex_app}
\end{equation}
Although the right-hand side involves complex matrices, the result is real symmetric:
$\sigma_{\hat{K}}=\sigma_{\hat{K}}^{\mathsf T}\in\mathbb R^{2n\times 2n}$.
It can be equivalently rewritten as:
\begin{equation}
\sigma_{\hat{K}}
=
\frac12\Big(\sigma_R+( \sigma_I + \Omega)\,\sigma_R^{-1}\,(\sigma_I +\Omega)^\tran \Big).
\label{eq:sigmaK_real_supp}
\end{equation}
It can also be shown explicitly that this is a bona fide covariance matrix of a quantum state, satisfying $\sigma_{\hat{K}} + \I \Omega \geq 0 $.

Finally, plugging Eqs.~\eqref{eq:p_supp} and~\eqref{eq:sigmaK_real_supp}
into Eq.~\eqref{eq:tracenorm_nu_supp} and finally in Eq.~\eqref{eq:tracenorm_mu_supp}, one obtains Eq.~\eqref{eq:tracenorm_mu} in the main text.
%

\section{Details on the OPO calculations}

Here we focus on the optical parametric oscillator (OPO), which we consider as a paradigmatic example to illustrate the applicability of our methods. In particular, after introducing its definition and its description within the Gaussian formalism, we present the details of the analytical calculation of the TSME-QFI for frequency estimation, the derivation of the SCGF together with the corresponding average current $\mathtt{J}$ and noise $\mathtt{D}$, as well as the TSME-QFI entering Hasegawa’s TUR.

The quantum-state dynamics of the OPO is described by the master equation
\begin{align}
    \dot{\rho} = -\I [\hat{H}_{\sf OPO},\rho] + \kappa\mathcal{D}[\hat{a}]\rho \,,
\end{align}
with $\hat{H}_{\sf OPO} = \omega \hat{a}^\dag \hat{a} + \I \frac{\chi}{2}\left( \hat{a}^2 - \hat{a}^{\dag 2} \right)$.
Within the Gaussian formalism, the evolution of the first-moment vector and covariance matrix of the state $\rho$ is governed by a Lyapunov equation with drift and diffusion matrices obtained from Eqs.~\eqref{eq:AandDdef},
\begin{align}
    A &=
         \begin{pmatrix}
                -\chi - \kappa/2 & \omega \\
               -\omega & \chi - \kappa/2 
        \end{pmatrix}\,, \\
    D &= \kappa \mathbb{1}_2 \,.
\end{align} 
By evaluating the eigenvalues of the drift matrix $A$, one finds that the system is stable, i.e., a steady state exists, if and only if $\chi^2 < \omega^2 + \kappa^2/4$.
In what follows, we restrict to this regime.

\subsubsection{Derivation of frequency estimation bounds}
\label{s:OPO_frequencyestimation}

To evaluate the ultimate bounds on frequency estimation via continuous monitoring, one first computes the matrices and vectors appearing in Eqs.~\eqref{eq:A12}–\eqref{eq:a_c_12}, obtaining
\begin{align}
    A_{\omega_1,\omega_2} &=
         \begin{pmatrix}
                -\chi - \frac{\kappa}2 & \frac{\omega_1+\omega_2}{2} \\
               -\frac{\omega_1+\omega_2}{2} & \chi - \frac{\kappa}{2} 
        \end{pmatrix}\,, \\
    D_{\omega_1,\omega_2} &= 
                \left[ \kappa + \I \left(\frac{\omega_1-\omega_2}{2}\right)\right] \mathbb{1}_2 \,, \\
    Q_{\omega_1,\omega_2} &= 
                - \I \left(\frac{\omega_1-\omega_2}{2}\right) \mathbb{1}_2 \,,
\end{align}
and $W_{\omega_1,\omega_2}=0$, $\mathbb{a}=\mathbb{c}=0$. The corresponding steady-state covariance matrix $\sigma_{\sf ss}$ and first-moment vector ${\bf d}_{\sf ss}$ are obtained by solving
\begin{align}
    0 &= \sigma_{\sf ss} Q_{\omega_1,\omega_2}\sigma_{\sf ss} + A_{\omega_1,\omega_2} \sigma_{\sf ss} + \sigma_{\sf ss}A_{\omega_1,\omega_2}^{T} + D_{\omega_1,\omega_2} \,, 
    \label{eq:steadysigma} \\
    0 &= A_{\omega_1,\omega_2} {\bf d}_{\sf ss} + \sigma_{\sf ss} Q_{\omega_1,\omega_2} {\bf d}_{\sf ss} - \sigma_{\sf ss}\Omega\mathbb{a}_{\omega_1,\omega_2} + \I \mathbb{c}_{\omega_1,\omega_2} \,.
    \label{eq:steadyd}
\end{align}

Remarkably, if we are interested in the parameter value $\omega=0$, and thus choose $\omega_1=0$ and $\omega_2=\varepsilon$, analytical solutions for both $\sigma_{\sf ss}$ and ${\bf d}_{\sf ss}$ can be obtained. The expression for $\sigma_{\sf ss}$ is cumbersome and not particularly informative, and is therefore omitted, while one straightforwardly finds ${\bf d}_{\sf ss}=0$. Substituting these results into Eq.~\eqref{eq:doteta}, one obtains the steady-state rate for the parameter $\xi$, namely
\begin{align}
    \dot{\xi}_{\sf ss} = - \vec{d}_{\sf ss}^\tran Q_{\omega_1,\omega_2} \vec{d}_{\sf ss} 
    - \frac{1}{2} \tr \left[ \sigma_{\sf ss} Q_{\omega_1,\omega_2} \right]
    - \frac{1}{2} \tr \left[ W_{\omega_1,\omega_2} \Omega \right]
    - 2 \mathbb{a}_{\omega_1,\omega_2}^\tran \Omega \vec{d}_{\sf ss} \,.
\end{align}
This expression can also be evaluated analytically, yielding a lengthy and not particularly illuminating function of $\kappa$, $\chi$, and $\varepsilon$.
However, the long-time TSME-QFI rate formula~\eqref{eq:QFIglobalGaussrate} simplifies considerably and leads to the analytical result
\begin{align}
\lim_{t \to \infty} \frac{\mathcal{Q} \left[ \ket{\Psi_{\mathrm{SE},\omega}(t)} \right]}{t} 
&=4 \Re\left[ \partial_\varepsilon^2 \dot\xi_{\sf ss}(\varepsilon) \bigl|_{\varepsilon=0} \right]  
=  \frac{8 \kappa  \chi ^2 \left(5 \kappa ^2-4 \chi ^2\right)}{\left(\kappa ^2-4 \chi ^2\right)^3} .
\label{eq:QFIrateOPO_SM}
\end{align}

At finite times, the time-dependent solutions for $\sigma(t)$, $\vec{d}(t)$, and $\xi(t)$ are obtained by integrating Eqs.~\eqref{eq:dotsigma}--\eqref{eq:doteta} and used to numerically evaluate the TSME-QFI at arbitrary times via Eq.~\eqref{eq:QFIglobalGauss}.
Likewise, they can be substituted into the expression for the environment-only fidelity~\eqref{eq:tracenorm_mu}, enabling the numerical evaluation of the environment-only QFI $\mathcal{Q}[ \rho_{\mathrm{E},\omega} ]$.
\subsubsection{Derivation of full counting statistics and Hasegawa's TUR bound}

The same procedure can be followed to derive the SCGF $C(\lambda)$ for quantum jumps. We first compute the matrices defined below Eq.~\eqref{eq:matrics_FCS}:
\begin{align}
        A_{\lambda} &=
         \begin{pmatrix}
                -\chi - \frac{\kappa e^{\I\lambda}}2 & \omega \\
               -\omega &  \chi - \frac{\kappa e^{\I\lambda}}2
        \end{pmatrix}\,, \\
    D_{\lambda} &= 
                \frac{\kappa}{2}\left(e^{\I \lambda} + 1\right)\mathbb{1}_2 \,, \\
    Q_{\lambda} &= 
               \frac{\kappa}{2}\left(e^{\I \lambda} - 1\right) \mathbb{1}_2\,,  \\
     W_{\lambda} &= 
        \begin{pmatrix}
              0 & \frac{\kappa}{2}\left(e^{\I \lambda} - 1\right) \\
             -\frac{\kappa}{2}\left(e^{\I \lambda} - 1\right) & 0 
        \end{pmatrix}\,,
\end{align}
with $\mathbb{a}_\lambda=\mathbb{c}_\lambda=0$. Proceeding as in Eqs.~\eqref{eq:steadysigma} and~\eqref{eq:steadyd}, we obtain the analytical steady-state solutions for the covariance matrix and first-moment vector. For $\omega=0$, one finds $\sigma_{ss}=\mathrm{diag}(s_{11},s_{22})$, with
\begin{align}
     s_{11} &= \frac{ e^{\I \lambda} \kappa + 2 \chi - \sqrt{\kappa^2 + 4 e^{\I \lambda} \kappa \chi +4\chi^2}}{\kappa ( e^{\I\lambda} -1)} \,,\\
    s_{22} &= \frac{ e^{\I \lambda} \kappa - 2 \chi - \sqrt{\kappa^2 - 4 e^{\I \lambda} \kappa \chi +4\chi^2}}{\kappa ( e^{\I\lambda} -1)} \,,
\end{align}
and ${\bf d}_{ss}=0$. Substituting these expressions into $\dot{\xi}_{ss}$ yields the SCGF
\begin{align}
    C(\lambda ) &:= \lim_{t\to \infty} \partial_t C(\lambda,t )= - \dot{\xi}_{\sf ss}(\lambda) \,,\\
                &=  \frac{1}{4} \biggl( 2\kappa -  \sqrt{\kappa ^2+4 \kappa  e^{ \I \lambda } \chi  +4 \chi ^2}  - \sqrt{\kappa ^2-4 \kappa  e^{ \I \lambda } \chi +4 \chi ^2} \biggr) \,,
\end{align}
from which the cumulants of the jump distribution can be directly evaluated, such as the average current, noise (variance rate), and skewness rate:
\begin{align}
    \mathtt{J} &= -\I \partial_{\lambda} C(\lambda) 
    = \frac{2 \kappa \chi^2 }{\kappa^2 - 4 \chi^2} \,, \\
    \mathtt{D} &= ( -\I \partial_{\lambda})^2 C(\lambda)
    = \frac{4 \kappa  \chi ^2 \left(\kappa ^4+2 \kappa ^2 \chi ^2+8 \chi ^4\right)}{\left(\kappa ^2-4 \chi ^2\right)^3} \,,\\
    \mathtt{S} &= ( -\I \partial_{\lambda})^3 C(\lambda)
    = \frac{8 \kappa \chi^2 \left(\kappa^8 + 14\kappa^6 \chi^2 + 84 \kappa^4 \chi^4 + 128 \kappa^2\chi^6 + 64 \chi^8\right)}{\left(\kappa ^2-4 \chi ^2\right)^5}\,.
\end{align}

As discussed in the main text, the ratio $\mathtt{D}/\mathtt{J}^2$ satisfies a TUR,
\begin{align}
    \frac{\mathtt{D}}{\mathtt{J}^2} \geq \frac{1}{f} \,,
\end{align}
where $f = \lim_{t \to \infty} \mathcal{Q}[ \ket{\Psi_{\mathrm{SE},\theta}(t)} ]/t$ is the joint TSME-QFI rate associated with a parameter $\theta$ that deforms the OPO dynamics according to $\omega \to \omega(1+\theta)$, $\chi \to \chi(1+\theta)$, and $\kappa \to \kappa(1+\theta)$. 

To evaluate $f$, one follows the same procedure described above for frequency estimation, now with respect to the parameter $\theta$. The matrices $A_{\theta_1,\theta_2}$, $D_{\theta_1,\theta_2}$, $Q_{\theta_1,\theta_2}$, and $W_{\theta_1,\theta_2}$ can be obtained from Eqs.~\eqref{eq:A12}–\eqref{eq:a_c_12}. Even restricting to $\omega=0$ and setting $\theta_1=0$, $\theta_2=\varepsilon$, these matrices are lengthy and not particularly informative, and we therefore omit them. In this case, a closed analytical expression for $\sigma_{ss}$ is not available. However, as follows from the TSME-QFI rate formula in Eq.~\eqref{eq:QFIrateOPO_SM}, only its expansion up to second order in $\varepsilon$ is required. By expanding Eqs.~\eqref{eq:steadysigma} and~\eqref{eq:steadyd} up to second order in $\varepsilon$, one can solve order by order and obtain the desired expansions for both $\sigma_{ss}$ and ${\bf d}_{ss}$. Although the corresponding expression for $\dot{\xi}_{ss}$ is again cumbersome and omitted, the resulting TSME-QFI can be evaluated analytically, yielding
\begin{align}
    f &= \lim_{t \to \infty} \frac{\mathcal{Q} \left[ \ket{\Psi_{\mathrm{SE},\theta}(t)} \right]}{t} 
    = 4 \Re\left[ \partial_\varepsilon^2 \dot\xi_{\sf ss}(\varepsilon) \bigl|_{\varepsilon=0} \right] 
    = \frac{2 \chi^2}{\kappa}\,.
\end{align}

\section{Finite-time photon counting statistics for an initial cat state}
\label{sec:cat_FCS}

We consider the Hamiltonian $\hat{H}=\omega\hat{a}^\dagger\hat{a}$ and jump operator $\hat{J}=\sqrt{\kappa}\hat{a}$, corresponding to the OPO with $\chi=0$, initialized in an even or odd cat state
\begin{equation}
\ket{\psi_\pm}
=
\frac{\ket{\alpha}\pm\ket{-\alpha}}
{\sqrt{2(1\pm e^{-2|\alpha|^2})}} \, .
\end{equation}
For photodetection efficiency $\eta$, the modified term in the tilted ME has coefficient $\kappa[1+\eta(e^{\I\lambda}-1)]$.

For the off-diagonal Gaussian operator $\hat{\mu}_{\beta\gamma}(0)=\ket{\beta}\bra{\gamma}$, Eqs.~\eqref{eq:dotsigma}--\eqref{eq:doteta} give no evolution for the covariance matrix $\sigma_{\beta\gamma}(t)=\id_2$ and
\begin{equation}
\Tr[\hat{\mu}_{\beta\gamma}(\lambda,t)]
=
\braket{\gamma|\beta}
\exp\!\left[\eta(1-e^{-\kappa t})(e^{\I\lambda}-1)\gamma^*\beta\right].
\end{equation}
Summing the traces of the four Gaussian operators obtained by evolving the initial state $\ket{\psi_\pm}\bra{\psi_\pm}$ yields
\begin{equation}
\varphi_+(\lambda,t) =\frac{\cosh\left[|\alpha|^2\,\left( 1+\eta(1-e^{-\kappa t})(e^{\I\lambda}-1)\right)\right]}{\cosh |\alpha|^2}, \quad
\varphi_-(\lambda,t) =\frac{\sinh\left[|\alpha|^2\,\left( 1+\eta(1-e^{-\kappa t})(e^{\I\lambda}-1)\right)\right]}{\sinh |\alpha|^2},
\end{equation}
In contrast, the incoherent equal mixture of
$\ket{\alpha}$ and $\ket{-\alpha}$ gives $\varphi_{\rm mix}(\lambda,t)
=\exp[|\alpha|^2 \eta(1-e^{-\kappa t}) (e^{\I\lambda}-1)]$.
For $\eta=1$ and complete decay, the cat-state counts retain their respective even or odd parity.
This is expected, since this evolution corresponds to time-resolved photon counting of the initial cat state.
%
\end{document}